\documentclass[12pt]{iopart}

\def\re{{\rm e}}
\def\rd{{\rm d}}
\def\N{{\mathbf N}}

\begin{document}

\title{$\lambda$-symmetries for discrete equations}
\author{
D  Levi$^{1}$ and   M A  Rodr\'{\i}guez$^{2}$}

\address{$^1$ Dipartimento di Ingegneria Elettronica \\ Universit\`a degli Studi Roma Tre and  INFN Sezione di Roma Tre\\ Via della Vasca Navale 84, 00146 Roma, Italy}
\address{$^2$ Departamento de F\'{\i}sica Te\'orica II\\ Facultad de F\'{\i}sicas, Universidad Complutense\\ 28040-Madrid, Spain}
\eads{\mailto{$^1$levi@roma3.infn.it}, \mailto{$^2$rodrigue@fis.ucm.es}}

\begin{abstract}
Following the usual definition of $\lambda$-symmetries of differential equations,  we introduce the analogous concept for difference equations and apply it to some examples.
\end{abstract}

\section{Introduction}
One of the most fruitful methods in the study of differential equations is the use of Lie symmetries to construct exact solutions. Once a symmetry is found, we can reduce the order of the ordinary differential equation  or, in the case of partial differential equations, construct special solutions as functions of the invariants of the symmetry group. 

As it is shown in Olver \cite{Ol95} there are  differential equations which can be reduced even if there is no Lie point symmetry. Muriel and Romero in 2001 \cite{MR01} introduced the concept of $\lambda$-symmetries  to  justify the existence of these special cases of reduction for ordinary differential equations.  Gaeta and Morando  gave later a geometrical interpretation for the $\lambda$-symmetries and extend it to partial differential equations ($\mu$-symmetries) \cite{GM04} (see also \cite {CM09,Ci08,CG07,Mo07,MR07,MR06} and \cite{Ga09} for a review of the problem). In these works, these symmetries were shown to be related to gauge transformations. Many other approaches to this problem have been proposed by different authors. For example, by Catalano Ferraioli \cite{Ca07}, using potential symmetries, or by Pucci and Saccomandi \cite{PS02}, using telescopic vector fields. 

Lie symmetry approach has been extended with success to the case of difference equations \cite{LW06}. From one side, we can discretize a differential equation with symmetries giving rise to a difference scheme, i.e. a set of difference equations defining both the equation and the lattice and from the other we can consider a discrete equation on a predefined lattice. In the first case the symmetries exist by construction and the purpose is to write and solve numerically the difference scheme. In the second case, when the equation and the lattice are a priori given, we would like to find solutions using symmetries but we are in a situation in which usually no symmetries are present. A way to get symmetries could be to use the approach to $\lambda$-symmetries. We show here that effectively we can construct $\lambda$-symmetries for discrete equations and we present some illustrative examples. 

 In Section 2 we review the Lie theory for $\lambda$-symmetries for ordinary differential equations. In Section 3 we present our definition of  $\lambda$-symmetries for ordinary difference equations together with some examples. Finally, Section 4 is devoted to some concluding remarks and to possible extensions.

\section{$\lambda$-symmetries for differential equations}
Let us briefly review the main ideas of the continuous $\lambda$-symmetry. 
If $\hat{X}$ is a vector field with variables $x$ and $u$,
\begin{equation} 
\hat{X}=\xi(x,u)\partial_x+\phi(x,u)\partial_u
\end{equation}
the coefficients of $\partial_{u_k}$, where $u_k=\frac{d^k u(x)}{d x^k}$, in the standard prolongation 
\begin{equation}\label{prol}
\hat{X}^{(m)}  = \xi(x,u) \partial_x +\phi(x,u) \partial_u +  \sum_{k=1}^m\phi^{(k)}  \partial_{u_{k}}
\end{equation}
are defined as \cite{Ol95}:
\begin{equation} \label{p1}
\phi^{(k+1)}=D_x\phi^{(k)}-u_{k+1}D_x\xi,\quad \phi^{(0)}=\phi,
\end{equation}
where by $D_x$ one means the total derivative with respect to $x$. 

The Lie symmetries of an $m$-th order ODE 
\begin{equation}\label{ode}
u_m = f(x,u,u_1,\ldots,u_{m-1})
\end{equation}
are obtained applying the $m$-prolonged infinitesimal generator (\ref{prol},\ref{p1}) to the equation, i.e., requiring that the invariance condition
\begin{equation}
\hat{X}^{(m)}(u_m-f)\big|_{u_m=f}=0
\end{equation}
be satisfied.

The $\lambda$-symmetries are defined as those symmetries for which the infinitesimal prolongation is modified with respect to the standard one (\ref{p1})  and is given by  \cite{MR01}:
\begin{equation}\label{muriel}
\phi^{(k+1,\lambda)}=(D_x+\lambda(x,u,u_1))\phi^{(k,\lambda)}-u_{k+1}(D_x+\lambda(x,u,u_1))\xi,
\end{equation}
where $\lambda(x,u,u_1)$ is a smooth function to be determined at the same time as the coefficients of the infinitesimal generators, $\xi$ and $\phi$. 
The $\lambda$-symmetry for an $m$-order differential equation is then obtained by applying  the following $m$-prolonged infinitesimal symmetry generator onto the differential equation:
\begin{equation}\label{bu3}
\hat{X}^{(m,\lambda)} = \xi \partial_x +\phi\, \partial_u +  \sum_{k=1}^m\phi^{(k,\lambda)}  \partial_{u_{k}}.
\end{equation}

If the ODE (\ref{ode}) is invariant under the symmetry generator $\hat X^{(m,\lambda)}$, then we say that it has a $\lambda$-symmetry. The $\lambda$-symmetries can be used to reduce the original equation and to find symmetry invariant solutions. 

As an example of ODE which has no Lie point symmetry but possesses $\lambda$-symmetries, let us consider the following ODE  (\cite{Ol95}, p. 182)
\begin{equation} \label{o1}
u_{2} = [ (x + x^2) \re^u]_x\, .
\end{equation}
This is an equation which can be integrated by quadrature but which has no symmetries. Equation (\ref{o1}) is written in the form of a conservation law,  
\begin{equation} \label{star}
u_{2}=D_x F(x,u)
\end{equation}
which has the obvious reduction  
\begin{equation}\label{red}
u_1 = F(x,u) + C.
\end{equation} 
Equation (\ref{star}) may  have no symmetries but we can prove that there always exists a $\lambda$-symmetry of infinitesimal generator given by $\xi=0$ and $\phi=1$ with $\lambda=F_u(x,u)$ which is at the origin of this reduction. The $\lambda$-symmetry generator is
\begin{equation}
\hat X^{(2,\lambda)} = \partial_u + F_u \partial_{u_1} + [ (F_u)^2 + u_1 F_{uu} + F_{xu} ] \partial_{u_2},
\end{equation}
which has two obvious invariants $z=x$ and $w=u_1-F$ and the differential invariant $w_z=D_x w/D_x z = u_2-F_x=0$. So the general solution of our equation is given by the solution of the equation (\ref{red}). For the particular choice of the function $F$ given by equation (\ref{o1}), the reduced equation is integrable and thus equation (\ref{o1}) is integrable by quadrature \cite{Ol95}.

This trivial example is not the only one we can find in the literature. For example, the following equation has been studied in \cite{Go88}
\begin{equation} \label{o3}
u_{2} - (u^{-1} (u_1)^2+ g(x) p u^p u_1 + g'(x) u^{p+1}) = 0.
\end{equation}
This equation is integrable by quadrature.  It has Lie point symmetries only when  the function $g(x)$ is given by one of the two following expressions,
\begin{equation} \label{o4}
g(x)=k_1 \re^{k_2 x} (k_3+ k_4 x)^{k_5}\quad \mbox{or}\quad g(x) = k_6 + \re^{k_7 x^2},
\end{equation}
where $k_j,\; j=1,\ldots, 7$ are arbitrary constant parameters. However, for a generic $g(x)$, there exists a $\lambda$-symmetry given by $\hat X= \partial_u$ and $\lambda=u^{-1}(u_1 + g(x) p u^{p+1})$.

How can we obtain the prolongation (\ref{muriel}) from the known theory? In what sense the $\lambda$-symmetries are a generalization of Lie point symmetries? These are some of the questions we need to answer to be able to construct $\lambda$-symmetries for equations on the lattice.

A constructive way to get $\lambda$-symmetries is contained in the work of Catalano Ferraioli \cite{Ca07}. There he introduces $\lambda$-symmetries exploiting  nonlocal symmetries. This approach is easily extendable to ordinary difference equations. The case of higher dimensional lattices  will be discussed elsewhere.

The construction presented in \cite{Ca07} consists fundamentally in adding to the ODE at study (\ref{ode}) an additional  differential equation 
\begin{equation}\label{aux}
w_1 = \lambda(x,u,u_1),
\end{equation}
for a new dependent function $w(x)$. This means that instead of considering an ODE we  are solving a system of ODEs written in triangular form. 
 One can state the following proposition (see \cite{Ca07} for the details):

{\bf Proposition:} {\it An ODE admits a $\lambda$-symmetry generator $\hat{X}^{(m,\lambda)}$ if and only if the symmetry generator 
\begin{equation}\label{gen}
\fl \hat{Y}^{(m)} = \xi(x,u,w) \partial_x +\phi(x,u,w) \partial_{u}+ \eta(x,u,u_1,w) \partial_{w}  +\sum_{i= 1}^m\phi^{(i)} \partial_{u_i} + \eta^{(1)}\partial_{w_1}
\end{equation}
 with 
\begin{equation}\label{total}
\eqalign{\phi^{(i)} = \tilde{D} \phi^{(i-1)} -  u_i\tilde{D} \xi,\quad \eta^{(1)} = \tilde{D} \eta  -  w_1\tilde{D} \xi,\\  \phi^{(0)}=\phi,\quad u_0=u,\quad w_0=w,}
\end{equation}
leaves the equations (\ref{ode}, \ref{aux}) invariant and is such that
\begin{equation}\label{com}
[\partial_w,\hat Y^{(m)}] = \hat Y^{(m)}.
\end{equation}}
In equation (\ref{total})  $\tilde D$ is the total derivative operator,
\begin{equation} 
\tilde{D} = \partial_x + \sum_{i\ge 0} u_{i+1} \partial_{u_i} + \sum_{i\ge 0} w_{i+1} \partial_{w_i}.
\end{equation}
Equation (\ref{com}) implies 
\begin{equation}\label{total1}
\eqalign{
\fl \xi(x,u,w) = \re^{w}\tilde{\xi}(x,u),\quad \phi\equiv\phi^{(0)}(x,u,w) =\re^{w} \tilde{\phi}(x,u),\quad \phi^{(i)}=\re^w\tilde{\phi}^{(i)},\nonumber \\ \fl  \eta^{(0)}  \equiv  \eta  (x,u,u_1,\ldots,u_{m-1},w)= \re^w \tilde{\eta}(x,u,u_1,\ldots,u_{m-1}),\quad \eta^{(1)}=\re^w\tilde{\eta}^{(1)},}
\end{equation}
and consequently 
\begin{equation} \label{24}
\hat Y^{(m)} = e^w \Big( \tilde \xi \partial_x + \sum_{i= 0}^m\tilde \phi^{(i)} \partial_{u_i} + \sum_{i=0}^1\tilde \eta^{(i)} \partial_{w_i} \Big).
\end{equation}
When we apply the generator (\ref{24}) onto the equation (\ref{aux})
\begin{equation}
\hat{Y}^{(1)}(w_1-\lambda(x,u,u_1))=e^w \big(\tilde \xi\partial_x+\tilde \phi\partial_{u}+\tilde \eta^{(1)}\partial_{w_1}\big)(w_1-\lambda(x,u,u_1))
\end{equation}
we get the determining equation for $\tilde{\eta}(x,u, u_1, \cdots, u_{m-1})$:
\begin{equation}\label{symcon}
\tilde \eta^{(1)}-\tilde \xi \lambda_x-\tilde \phi \,\lambda_u-\tilde{\phi}^{(1)}\lambda_{u_1}=0
\end{equation}
where, from (\ref{total}, \ref{total1})
\begin{equation}
\fl\eqalign{\tilde{\eta}^{(1)}=\re^{-w}(\tilde{D}_x\eta-w_1\tilde{D}_x\xi)=\tilde{\eta}_x+w_1\tilde{\eta}+\sum_{k=1}^m u_k\tilde{\eta}_{u_{k-1}} -(\tilde{\xi}_x+w_1\tilde{\xi}+u_1\tilde{\xi}_u)w_1\\
\tilde{\phi}^{(1)}=\re^{-w}(\tilde{D}_x\phi-u_1\tilde{D}_x\xi)=\tilde{\phi}_x+w_1\tilde{\phi} -(\tilde{\xi}_x+w_1\tilde{\xi}- \tilde{\phi}_{u}+u_1\tilde{\xi}_u)u_1}
\end{equation}
Applying the generator $\hat{Y}^{(m)}$ (\ref{24}) onto (\ref{ode}), we can factorize the $w$-dependence and the infinitesimal generator $\hat{Y}^{(m)}$ reduces to the $\lambda$-symmetry generator (\ref{bu3}). When equation (\ref{aux}) is satisfied, equation (\ref{symcon}) is a $w$-independent partial differential equation for $\tilde \eta$ in terms of $\tilde{\phi}$ and $\lambda$:
\begin{eqnarray}\label{30}
\tilde{\eta}_x+\sum_{k=1}^{m-1} u_k\tilde{\eta}_{u_{k-1}}+f\tilde{\eta}_{u_{m-1}}=\tilde{\xi}\lambda_x+\tilde{\phi}\lambda_u \nonumber\\  +(\tilde{\phi}_x+\lambda\tilde{\phi}-(\tilde{\xi}_x+\lambda\tilde{\xi}-\tilde{\phi}_{u}+u_1\tilde{\xi}_u)u_1)\lambda_{u_1}+(\tilde{\xi}_x+\lambda\tilde{\xi}+u_1\tilde{\xi}_u-\tilde{\eta})\lambda\qquad
\end{eqnarray}

In this way, $\lambda$ symmetries are just classical symmetries for the system formed by the ODE (\ref{ode}) and the equation (\ref{aux}). However, they may not correspond to just Lie point symmetries as equation (\ref{30}) may not have a solution when $\tilde \eta$ depends just on $x$ and $u$. See \cite{Ca07} for more details on the equivalence of this method to that of Muriel and Romero \cite{MR01}.

\section{$\lambda$-symmetries in the discrete case}

We will construct $\lambda$-symmetries in the discrete case, closely following the approach we have discussed in the last section. Let us consider a difference scheme in a one-dimensional lattice:
\begin{equation}\label{disc}
f_i(x_{n-a},\ldots,x_{n+b},u_{n-a},\ldots,u_{n+b})=0,\quad a,b\in \N,\quad i=1,2
\end{equation}
Equations (\ref{disc}) correspond to a discrete scheme where the two equations define at the same time the solution and the lattice. In the continuous limit when the distance between the points goes to zero, one of the equations is identically satisfied while the other goes over to a differential equation \cite{LW06}.

In order to find $\lambda$-symmetries for this equation, we introduce a first order difference equation for a new dependent variable $w_n$:
\begin{equation}\label{disaux}
w_{n+1}-w_n-(x_{n+1}-x_n)\lambda_n(x_n,u_n)=0
\end{equation}
Here, we show explicitly the $\lambda_n$ dependence on the values of  $u_n$ in the  point of the lattice of index $n$ but we must think that $\lambda_n$ may depend on more lattice points, i.e., $\lambda_n(x_n,\{u_j\}_{j=-\alpha}^{j=\beta})$, $\alpha,\beta\in \mathbf{N}$.
From equation (\ref{disaux}) we can express the function $w_n$ in any point of the lattice in terms of the function $u_n$ and the initial data. In fact, by solving it we get:
\begin{equation} \label{32}
\eqalign{w_{n+k} = w_n +  \sum_{i=0}^{k-1}(x_{n+i+1}-x_{n+i})\lambda_{n+i}(x_{n+i},u_{n+i}), \quad k \in \N  \\ 
w_{n-k} = w_n -  \sum_{i=0}^{k-1} (x_{n-i}-x_{n-i-1})\lambda_{n-i-1}(x_{n-i-1},u_{n-i-1}), \quad k \in \N.}
\end{equation}
 The symmetry generator for equations (\ref{disc}, \ref{disaux}) is:
\begin{equation} \label{yd}
\hat{Y} = \xi_{n}(x_{n},u_{n},w_{n})\partial_{x_{n}}+\phi_{n}(x_{n},u_{n},w_{n})\partial_{u_{n}}  +\eta_{n}(x_{n},u_{n},w_{n})\partial_{w_{n}} 
\end{equation}
where, as in the case of the function $\lambda_n$, $\eta_n$ may depend on more lattice points. The prolongation of $\hat{Y}$ is given by
\begin{eqnarray}
\fl  \hat{Y}^{(a,b)} = & \sum_{i=-a}^b\xi_{n+i}(x_{n+i},u_{n+i},w_{n+i})\partial_{x_{n+i}}+\sum_{i=-a}^b\phi_{n+i}(x_{n+i},u_{n+i},w_{n+i})\partial_{u_{n+i}}+ \nonumber \\ \fl & +\sum_{i=0}^1\eta_{n+i}(x_{n+i},u_{n+i},w_{n+i})\partial_{w_{n+i}}.
\end{eqnarray}
If we apply $ \hat{Y}^{(0,1)}$ onto equation (\ref{disaux}) we get,
\begin{equation}\label{discinv}
-(\xi_{n+1}-\xi_{n})\lambda_n+(x_{n+1}-x_{n})(\xi_{n}\partial_{x_{n}}\lambda_n+\phi_{n}\partial_{u_{n}}\lambda_n)+\eta_{n+1}-\eta_n=0.
\end{equation}
For simplicity we consider here that $\lambda_n=\lambda_n(x_n,u_n)$ only, but the result, with appropriated changes, is valid in general.

Imposing the condition (\ref{com}) to the symmetry generator (\ref{yd}) we get
\begin{equation}\label{sym}
\eqalign{&\xi_{n+i} = \re^{w_{n+i}}\tilde{\xi}_{n+i}(x_{n+i},u_{n+i}),\quad \phi_{n+i} =\re^{w_{n+i}}\tilde{\phi}_{n+i}(x_{n+i},u_{n+i}),\nonumber\\ & \eta_{n+i} =\re^{w_{n+i}}\tilde{\eta}_{n+i}(x_{n+i},u_{n+i}).}
\end{equation}
If we substitute (\ref{sym}) in (\ref{discinv}), we obtain
\begin{equation}
\fl \re^{w_{n+1}-w_n}(\tilde{\eta}_{n+1}-\tilde{\xi}_{n+1}\lambda_n)-(\tilde{\eta}_n-\tilde{\xi}_{n} \lambda_n)+ (x_{n+1}-x_{n})( \tilde{\xi}_{n}\partial_{x_{n}}\lambda_n + \tilde{\phi}_{n}\partial_{u_{n}}\lambda_n) =0,
\end{equation}
and using (\ref{disaux}), we find that $\tilde{\eta}_n$ must satisfy the following $w_n$-independent equation
\begin{equation}\label{condy}
\fl \re^{(x_{n+1}-x_n)\lambda_n}(\tilde{\eta}_{n+1}-\tilde{\xi}_{n+1}\lambda_n)-(\tilde{\eta}_n-\tilde{\xi}_{n} \lambda_n)   + (x_{n+1}-x_{n})( \tilde{\xi}_{n}\partial_{x_{n}}\lambda_n + \tilde{\phi}_{n}\partial_{u_{n}}\lambda_n)  =  0
\end{equation}

It is worthwhile to notice that equation (\ref{condy}) implies that the function $\tilde{\eta}_n$ will depend in general on several points of the lattice; its solution when $\tilde \eta_n$ depends just on $u_n$ maybe trivial as equation (\ref{condy}) imposes in this case strict constraints on $\lambda_n$, $\tilde \xi_n$ and $\tilde \phi_n$.

The prolongation of the infinitesimal generator of the $\lambda$-symmetry is given by
\begin{equation}\label{prold}
\fl \hat{X}^{(a,b;\lambda)}=\tilde{\xi}_{n} \partial_{x_{n+k}}+\tilde{\phi}_{n} \partial_{u_{n}}+\sum_{k=1}^{b} \re^{w_{n+k}-w_n}\tilde{\phi}_{n+k} \partial_{u_{n+k}}+\sum_{k=1}^{a} \re^{w_{n-k}-w_n}  \tilde{\phi}_{n-k} \partial_{u_{n-k}},
\end{equation}
and, when we apply (\ref{prold}) onto equations (\ref{disc}), we get
\begin{equation}
\fl \sum_{k=-a}^{b} \re^{w_{n+k}} (\tilde{\xi}_{n+k} \partial_{x_{n+k}}f_i+\tilde{\phi}_{n+k} \partial_{u_{n+k}}f_i) \big|_{[f_i= 0, \; w_{n+1}=w_n+\lambda_n(x_{n+1}-x_n)]} = 0,\quad i=1,2,
\end{equation}
which, taking into account equations (\ref{32}), provide a factorized common factor $\re^{w_n}$ in front of an equation depending only on $u_n$. So, as the dependence on $w_n$ factorizes, the effective infinitesimal generator (\ref{prold}) when applied to the O$\Delta$E (\ref{disc}) will depend just on $\xi_n$, $\phi_n$ and $\lambda_n$. In this way we get the extra freedom necessary to possibly get nontrivial symmetries.

To check the correspondence of this approach with the continuous case in \cite{MR01}, let us consider the two-point prolongation
\begin{equation}\label{vectfield}
\hat{X}^{(0,1;\lambda)}=  \tilde{\xi}_{n}\partial_{x_{n}}+ \tilde{\phi}_{n}\partial_{u_{n}}  +\re^{(x_{n+1}-x_n)\lambda_n}\big(\tilde{\xi}_{n+1}\partial_{x_{n+1}}+\tilde{\phi}_{n+1}\partial_{u_{n+1}}\big) .
\end{equation}
In this case we can construct the approximation to the first derivative and consequently we can get the continuous limit formula corresponding to the first prolongation (equation (\ref{bu3}) with $m=1$).

Let us consider the following change of variables, from  $(x_n,x_{n+1},u_n,u_{n+1})$ to $(\bar{x}_n,h_{n+1},\bar{u}_{n},u_{x,n})$:
\begin{equation}
\bar{x}_n=x_n,\quad \bar{u}_n=u_n,\quad h_{n+1}=x_{n+1}-x_n,\quad u_{x,n+1}=\frac{u_{n+1}-u_n}{x_{n+1}-x_n} 
\end{equation}
which corresponds to the infinitesimal transformation:
\begin{equation}
\eqalign{\partial_{x_n}= \partial_{\bar{x}_n}+\frac{u_{x,n+1}}{h_{n+1}}\partial_{u_{x,n+1}}- \partial _{h_{n+1}}, 
\quad  
\partial_{x_{n+1}}= -\frac{u_{x,n+1}}{h_{n+1}}\partial_{u_{x,n+1}}+\partial _{h_{n+1}}  \\   
\partial_{u_n}= \partial_{\bar{u}_n}-\frac{1}{h_{n+1}}\partial_{u_{x,n+1}}, 
\quad
\partial_{u_{n+1}}=  \frac{1}{h_{n+1}}\partial_{u_{x,n+1}}.}
\end{equation}
Then the vector field (\ref{vectfield}) is rewritten as:
\begin{eqnarray}
\hat{X}^{(0,1;\lambda)}  = &
\tilde{\xi}_{n}\partial_{\bar{x}_n}
+ \tilde{\phi}_{n}\partial_{\bar{u}_n}
+(\re^{h_{n+1}\lambda_n}\tilde{\xi}_{n+1}-\tilde{\xi}_{n})\partial _{h_{n+1}}
\nonumber \\ & 
+\bigg(
\frac{\re^{h_{n+1}\lambda_n}\tilde{\phi}_{n+1}-\tilde{\phi}_{n}} {h_{n+1}}
-\frac{\re^{h_{n+1}\lambda_n}\tilde{\xi}_{n+1}-\tilde{\xi}_{n}}{h_{n+1}}
u_{x,n+1} \bigg)
\partial_{u_{x,n+1}}.
\end{eqnarray}
In the continuous limit, $h_n$ and  $h_{n+1}$ go to zero, so that we have take $\re^{h_{n+1}\lambda_{n}}\sim 1+h_{n+1}\lambda_{n}$, and $u_{x,n+1}$ goes into $u_1$. Then
\begin{eqnarray}
\fl \hat{X}^{(0,1;\lambda)}  \sim &
\lim_{h\to 0}\bigg(\tilde{\xi}_{n}\partial_{\bar{x}_n}
+ \tilde{\phi}_{n}\partial_{\bar{u}_n}
+\big(\tilde{\xi}_{n+1}-\tilde{\xi}_{n}+h_{n+1}\lambda_n\tilde{\xi}_{n+1}\big)\partial _{h_{n+1}}
\nonumber \\ \fl  &
+\bigg(
\frac{\tilde{\phi}_{n+1}-\tilde{\phi}_{n}} {h_{n+1}}
-\frac{\tilde{\xi}_{n+1}-\tilde{\xi}_{n}}{h_{n+1}}
u_{x,n+1}
+\lambda_n(\tilde{\phi}_{n+1}-\tilde{\xi}_{n+1}u_{x,n+1})
\bigg)
\partial_{u_{x,n+1}}\bigg)\nonumber \\ 
\fl &  
=\tilde{\xi}\,\partial_{x}
+ \tilde{\phi}\,\partial_{u} 
+\big[(D_x+\lambda)\tilde{\phi}
-u_1(D_x+\lambda)\tilde{\xi}
\big]
\partial_{u_1}
\end{eqnarray}
i.e., the result of Muriel and Romero \cite{MR01}.

\subsection{Examples} Let us consider as examples, second order difference equations on a fixed untransformable lattice of spacing $x_{n+1}-x_n=h$ which  has $\lambda$-symmetries:
\begin{equation}\label{sec}
\frac{u_{n+1}-2 u_n + u_{n-1} }{h^2}=F_n(u_n,u_{n-1})
\end{equation}
Taking into account the results presented above, the determining equation for the $\lambda$-symmetries of equation (\ref{sec}) is obtained by applying the vector field
\begin{equation}
\hat{X}^{(1,1;\lambda)}=\tilde{\phi}_n\partial_{u_n}+\re^{h\lambda_n(u_n)}\tilde{\phi}_{n+1}\partial_{u_{n+1}}+\re^{-h\lambda_{n-1}(u_{n-1})}\tilde{\phi}_{n-1}\partial_{u_{n-1}}
\end{equation}
onto the equation (\ref{sec}). If we define $\chi_n=\re^{h\lambda_n(u_n)}$, the vector field reduces to
\begin{equation}\label{vec}
\hat{X}^{(1,1;\lambda)}=\tilde{\phi}_n\partial_{u_n}+\chi_n\tilde{\phi}_{n+1}\partial_{u_{n+1}}+\frac{\tilde{\phi}_{n-1}}{\chi_{n-1}}\partial_{u_{n-1}}
\end{equation}
and the determining equation is
\begin{equation}\label{det}
\chi_n\tilde{\phi}_{n+1}+\frac{\tilde{\phi}_{n-1}}{\chi_{n-1}}-2\tilde{\phi}_n=h^2\bigg(\tilde{\phi}_nF_{n,u_n}+\frac{\tilde{\phi}_{n-1}}{\chi_{n-1}}F_{n,u_{n-1}}\bigg).
\end{equation}

{\bf Example 1.}
Let us choose a function $F_n$ which is the discrete derivative of a function $f_n(u_n)$, i.e.,
\begin{equation}\label{ex1}
\frac{u_{n+1}-2 u_n + u_{n-1} }{h^2}=F_n(u_n,u_{n-1})=\frac{1}{h}[f_n(u_n)-f_{n-1}(u_{n-1})].
\end{equation}
This equation has no point symmetry for a generic function $f_n(u_n)$. 
It is obvious that this equation, as in the example of Olver (\ref{star}), reduces to a first order difference equation
\begin{equation}\label{first1}
\frac{u_{n+1}-u_n}{h}=f_n(u_n)+C.
\end{equation}
We will show that it has a non trivial $\lambda$-symmetry which will be the origin of the reduction of the order of the equation. If we take $\tilde{\phi}_n=1$ the determining equation (\ref{det}) reduces to
\begin{equation}
\chi_n+\frac{1}{\chi_{n-1}}-2=h\bigg(\partial_{u_{n}}f_{n}-\frac{1}{\chi_{n-1}}\partial_{u_{n-1}}f_{n-1}\bigg)
\end{equation}
and is satisfied by $\chi_n=1+h \partial_{u_n}f_{n}$. The vector field (\ref{vec}) is in this case
\begin{equation}
\hat{X}^{(1,1;\lambda)}=\partial_{u_n}+(1+h\partial_{u_n}f_n)\partial_{u_{n+1}}+ \frac{1}{1+h\partial_{u_{n-1}}f_{n-1}} \partial_{u_{n-1}}
\end{equation}
and its invariant is 
\begin{equation}
v_n=u_{n+1}-u_n-hf_n(u_n),
\end{equation}
Substituting into equation (\ref{ex1}) we get $v_n=C$, i.e., $u_n$ satisfies  (\ref{first1}).
\bigskip

{\bf Example 2.}
Let us consider the difference equation
\begin{equation}\label{ex}
\frac{u_{n+1}-2 u_n + u_{n-1} }{h^2}= u_{n-1}\bigg(1 + \frac{h}{2} u_{n-1}\bigg)\frac{u_n - u_{n-1}}{h}- \frac{h}{8}  u_{n-1}^4,
\end{equation}
whose continuous limit gives the ODE 
\begin{equation}
u_2 = u u_1 +\frac{1}{2}\bigg( u^2 u_1 - u u_2 -2 u_1^2 - \frac{1}{4} u^4\bigg) h 
+\mathcal O(h^2).
\end{equation}
Equation (\ref{ex}), being on an uniform lattice, could only have dilation symmetries. It is easy to see that it has no Lie  symmetry. Taking into account the results presented above, the determining equation for the $\lambda$-symmetries of equation (\ref{ex}) is given by
\begin{eqnarray}\label{dex}
\chi_n \tilde{\phi}_{n+1}-2 \tilde{\phi}_n+\frac{1}{\chi_{n-1}}\tilde{\phi}_{n-1}=\left(1+\frac{1}{2} h u_{n-1}\right)h u_{n-1} \tilde{\phi}_n+\nonumber \\ \qquad h\left(u_n-2 u_{n-1} + h u_{n-1} \left(u_n-\frac{3}{2} u_{n-1}\right)-\frac{1}{2} h^2u_{n-1}^3\right) \frac{1}{\chi_{n-1}} \tilde{\phi}_{n-1}
\end{eqnarray}
when equation (\ref{ex}) is satisfied, i.e. when $u_{n+1}$ is expressed in terms of $u_n$ and $u_{n-1}$ using this equation. To get a $\lambda$-symmetry we choose $\tilde \phi_n=1$ and equation (\ref{dex}) reduces to just a nonlinear first difference equation for the function $\chi_n(u_n)$
\begin{equation} \label{ddex}
\chi_n - 2 (1 + \frac{1}{2} h u_{n-1})^2 + \frac{(1+h u_{n-1})[(1+h u_{n-1})^2 +1 - 2 h u_n ] }{2 \chi_{n-1}} = 0.
\end{equation}
If we differentiate equation (\ref{ddex}) with respect to $u_n$ twice we get $\ddot \chi_n=0$, i.e., $\chi_n = \alpha + \beta u_n$ where $\alpha$ and $\beta$ are two arbitrary constants. Substituting this necessary result into equation (\ref{ddex}) we get that $\alpha=1$, $\beta=h$ and taking into account the definition of $\chi_n$ we get  as the only possible solution
\begin{equation}\label{lam}
\lambda_n =\frac{1}{h} \log(1+h u_n)
\end{equation}

We can now use the $\lambda$-symmetry to reduce the difference equation. The infinitesimal generator of the $\lambda$-symmetry is
\begin{equation} \nonumber
\hat{X}^{(1,1;\lambda)} = \frac{1}{\chi_{n-1}}\partial_{u_{n-1}}+ \partial_{u_n}+\chi_{n} \partial_{u_{n+1}}
\end{equation}
The invariants are obtained by integrating the following shift related equations
\begin{equation}\label{equ}
\chi_{n-1}\rd u_{n-1}=\frac{\rd u_n}{1} = \frac{\rd u_{n+1}}{\chi_{n}}.
\end{equation}
which, after substituting  (\ref{lam}), yields
\begin{equation}\label{equ2}
(1 + h u_{n-1}) \rd u_{n-1}=\rd u_n = \frac{\rd u_{n+1}}{1 + h u_n}.
\end{equation}
The compatible solution of equations (\ref{equ2}) is given by
\begin{equation}\label{comp}
u_{n+1} -u_n = \kappa_n + \frac{h}{2} u_{n}^2
\end{equation}
where $\kappa_n$ appear as an integration constant and thus depends just on $n$, the invariant index. If we introduce equation (\ref{comp}) into the nonlinear equation (\ref{ex}) we get as the reduced equation, a logistic--type difference equation
\begin{equation}
\kappa_{n+1} = \kappa_{n} \bigg(1 - \frac{h}{2} \kappa_{n}\bigg).
\end{equation}
This is a first order recurrence relation. One could look again for a $\lambda$-symmetry. In this case one can always find one but we can not use it to get a solution.

\section{Conclusions}
In this paper we have considered the $\lambda$-symmetries for difference equations. They are determined by extending to the discrete case the potential symmetries in the form given by Catalano Ferraioli.

As one can see from equation (\ref{prold}) the discrete $\lambda$-prolongation is quite different form the continuous case (\ref{muriel}, \ref{bu3}) and it would be difficult to guess it without going through the potential symmetries.

The examples we presented, show that also in the discrete case the $\lambda$-symmetries can be used to reduce the equations to simpler ones as in the continuous case. It is worthwhile to notice that discrete equations have usually much less symmetries that the continuous one, so $\lambda$-symmetries can be in this case more useful.

Work is in progress to extend this result to the case of $\mu$-symmetries, i.e., to the case of partial difference equations.

\ack DL has been partly supported by the Italian Ministry of Education and Research, PRIN ``Nonlinear waves: integrable fine dimensional reductions and discretizations'' from 2007 to 2009 and PRIN ``Continuous and discrete nonlinear integrable evolutions: from water waves to symplectic maps'' from 2010. The work of MAR has been partly supported by MCI (Spain), grant FIS2008-00209, and UCM-Banco Santander, grant GR58/08-910556. We are greatly indebted to G. Gaeta for suggesting us this problem and to C. Muriel for enlightening discussions on this subject.


\section*{References}

\begin{thebibliography}{99}
\bibitem{Ca07} Catalano Ferraioli D 2007 Nonlocal aspects of $\lambda$-symmetries and ODEs reduction, {\it J. Phys. A: Math. Theor.} {\bf 40}  5479--5489

\bibitem{CM09} Catalano Ferraioli D and  Morando P 2009 Local and nonlocal solvable structures in ODEs reduction, {\it J. Phys. A: Math. Theor.} {\bf 42}  035210

\bibitem{Ci08} Cicogna G 2008 Reduction of systems of first-order differential equations via $\lambda$-symmetries, {\it Phys. Lett.} A {\bf 372}  3672--3677

\bibitem{CG07} Cicogna G and Gaeta G 2007 Noether theorem for $\mu$-symmetries, {\it J. Phys. A: Math. Theor.} {\bf 40}  11899--11921

\bibitem{Ga09} Gaeta G 2009 A gauge-theoretic description of $\mu$-prolongations, and $\mu$-symmetries of differential equations, {\it J. Geom. Phys.} {\bf 59}  519--539

\bibitem{GM04} Gaeta G and Morando P 2004 On the geometry of $\lambda$-symmetries and PDEs reduction, {\it J. Phys. A: Math. Gen.} {\bf 37}  6955--6975

\bibitem{Go88} Gonz\'alez-L\'opez A 1988 Symmetry  and integrability by quadratures of ordinary differential equations, {\it Phys. Lett.} A {\bf 133} 190--194

\bibitem{LW06} Levi D and Winternitz P 2006  Continuous symmetries of difference equations, {\it J. Phys. A: Math. Gen.} {\bf 39}  R1--R36

\bibitem{Mo07} Morando P 2007 Deformation of Lie derivative and $\lambda$-symmetries, {\it J. Phys. A: Math. Theor.} {\bf 40} 11547--11559

\bibitem{MR01} Muriel C and Romero J L 2001 New method of reduction for ordinary differential equations, {\it IMA J. Appl. Math.} {\bf 66} 111--125

\bibitem{MR07} Muriel C and Romero J L  2007 $C^{\infty}$-symmetries and nonlocal symmetries of exponential type, {\it IMA J. Appl. Math.} {\bf 72} 191--205

\bibitem{MR06} Muriel C, Romero J L and Olver P J 2006 Variational $C^{\infty}$ symmetries and Euler-Lagrange equations, {\it J. Diff. Eqs.} {\bf 222} 164--184

\bibitem{Ol95} Olver P J  1995 {\it Applications of Lie groups to differential equations} (New York: Springer)

\bibitem{PS02} Pucci E and Saccomandi G 2002 On the reduction methods for ordinary differential equations, {\it J. Phys. A: Math. Gen.} {\bf 35}  6145--6155
\end{thebibliography}
\end{document}